\documentclass[12pt]{article}
\usepackage{epsf}
\usepackage{graphicx}

\begin{document}
\title
{Low-energy quantum gravity}
\author
{Michael A. Ivanov \\
Physics Dept.,\\
Belarus State University of Informatics and Radioelectronics, \\
6 P. Brovka Street,  BY 220027, Minsk, Republic of Belarus.\\
E-mail: ivanovma@gw.bsuir.unibel.by.}

\maketitle

\begin{abstract}If gravitons are super-strong interacting particles and the
low-temperature graviton background exists, the basic cosmological
conjecture about the Dopplerian nature of redshifts may be false.
In this case, a full magnitude of cosmological redshift would be
caused by interactions of photons with gravitons. Non-forehead
collisions with gravitons will lead to a very specific additional
relaxation of any photonic flux. It gives a possibility of another
interpretation of supernovae 1a data - without any kinematics.
These facts may implicate a necessity to change the standard
cosmological paradigm.
\par A quantum mechanism of classical gravity based on an
existence of this sea of gravitons is described for the Newtonian
limit. This mechanism needs graviton pairing and "an atomic
structure" of matter for working it, and leads to the time
asymmetry. If the considered quantum mechanism of classical
gravity is realized in the nature, then an existence of black
holes contradicts to Einstein's equivalence principle. It is shown
that in this approach the two fundamental constants - Hubble's and
Newton's ones - should be connected between themselves. The
theoretical value of the Hubble constant is computed. In this
approach, every massive body would be decelerated due to
collisions with gravitons that may be connected with the Pioneer
10 anomaly. Some unsolved problems are discussed, so as
possibilities to verify some conjectures in laser-based
experiments.
\end{abstract}
\section[1]{Introduction }
An opinion is commonly accepted that quantum gravity should
manifest itself only on the Planck scale of energies, i.e. it is a
high-energy phenomenon. The value of the Planck energy $\sim
10^{19}$ GeV has been got from dimensional reasonings. In this
contribution, I would like to describe a very unexpected
possibility to consider gravity as a very-low-energy stochastic
process. I enumerate those discoveries and observations which may
support this my opinion.
\par 1. In 1998, Anderson's team reported about the discovery of
anomalous acceleration of NASA's probes Pioneer 10/11 \cite{1};
this effect is not embedded in a frame of the general relativity,
and its magnitude is somehow equal to $\sim Hc$, where $H$ is the
Hubble constant, $c$ is the light velocity. \par 2. In the same
1998, two teams of astrophysicists, which were collecting
supernovae 1a data with the aim to specificate parameters of
cosmological expansion, reported about dimming remote supernovae
\cite{2,3}; the one would be explained on a basis of the Doppler
effect if at present epoch the universe expands with acceleration.
This explanation needs an introduction of some "dark energy" which
is unknown from any laboratory experiment. \par 3. In January
2002, Nesvizhevsky's team reported about discovery of quantum
states of ultra-cold neutrons in the Earth's gravitational field
\cite{4}. Observed energies of levels (it means that and their
differences too) in full agreement with quantum-mechanical
calculations turned out to be equal to $\sim 10^{-12}$ eV. If
transitions between these levels are accompanied with irradiation
of gravitons then energies of irradiated gravitons should have the
same order - but it is of 40 orders lesser than the Planck energy.
\par An alternative model of redshifts \cite{5} which
is based on a conjecture about an existence of the graviton
background gives us odds to see on the effect of supernova dimming
as an additional manifestation of low-energy quantum gravity. The
main results of author's research in this approach are described
here briefly (it is a short version of my summarizing paper
\cite{500}). Starting from a statistical model of the graviton
background with a low temperature, it is shown - under the very
important condition that gravitons are super-strong interacting
particles - that if a redshift would be a quantum gravitational
effect then one can get from its magnitude an estimate of a new
dimensional constant characterizing a single act of interaction in
this model.

\section[2]{Passing photons through the graviton background \cite{5,5a}}

If the isotropic graviton background exists, then it is possible
photon scattering on gravitons, if one of the gravitons is
virtual. Due to forehead collisions with gravitons, an energy of
any photon should decrease when it passes through the sea of
gravitons. From another side, none-forehead collisions of photons
with gravitons of the background will lead to an additional
relaxation of a photon flux, caused by transmission of a momentum
transversal component to some photons. It will lead to an
additional dimming of any remote objects, and may be connected
with supernova dimming. We deal here with the uniform
non-expanding universe with the Euclidean space, and there are not
any cosmological kinematic effects in this model. We shall take
into account that a gravitational "charge" of a photon must be
proportional to $E$ (it gives the factor  $E^{2}$ in a
cross-section) and a normalization of a photon wave function gives
the factor $E^{-1}$ in the cross-section. Also we assume here that
a photon average energy loss $\bar \epsilon $ in one act of
interaction is relatively small to a photon energy $E.$ Then
average energy losses of a photon with an energy  $E $ on a way
$dr $ will be equal to \cite{5,5a}:
\begin{equation}
                  dE=-aE dr,
\end{equation}
where $a$ is a constant. If a {\it whole} redshift magnitude is
caused by this effect, we must identify $a=H/c,$ where $c$ is the
light velocity, to have the Hubble law for small distances. The
expression (1) is true if the condition $\bar \epsilon << E(r)$
takes place. Photons with a very small energy may lose or acquire
an energy changing their direction of propagation after
scattering.  Early or late such photons should turn out in the
thermodynamic equilibrium with the graviton background, flowing
into their own background. Perhaps, the last one is the cosmic
microwave background.

\par Photon flux's average energy losses on a way $dr$ due to
non-forehead collisions with gravitons should be proportional to
$badr,$ where $b$ is a new constant of the order $1.$ These losses
are connected with a rejection of a part of photons from a
source-observer direction. We get for the factor $b$ (see
\cite{213}):
\begin{equation}
b=\frac {4}{\pi} (\int_{0}^{\pi/4}2\cos^{2}{\theta}d\theta +
\int_{\pi/4}^{\pi/2}\sin^{2}{2\theta}d\theta)= \frac {3}{2} +
\frac {2}{\pi} \simeq 2.137.
\end{equation}
\par
Both redshifts and the additional relaxation of any photonic flux
due to non-forehead collisions of gravitons with photons lead in
our model to the following luminosity distance $D_{L}:$
\begin{equation}
D_{L}=a^{-1} \ln(1+z)\cdot (1+z)^{(1+b)/2} \equiv a^{-1}f_{1}(z),
\end{equation}
where $f_{1}(z)\equiv \ln(1+z)\cdot (1+z)^{(1+b)/2}$. \par To
compare a form of this predicted dependence $D_{L}(z)$ by unknown,
but constant $H$, with the latest observational supernova data by
Riess et al. \cite{203}, we can use the fact that $f_{1}$ is the
luminosity distance in units of $c/H$. In Figure 1, the graph of
$f_{1}$ is shown; observational data (82 points) are taken from
Table 5 of \cite{203}. Observations of \cite{203} are transformed
as $\mu_{0} \rightarrow 10^{(\mu_{0}-c_{1})/5}$ with the constant
$c_{1}=43$. The predictions fit observations very well for roughly
$z < 0.5$. It excludes a need of any dark energy to explain
supernova dimming. Discrepancies between predicted and observed
values of $\mu_{0}(z)$ are obvious for higher $z$: we see that
observations show brighter SNe that the theory allows, and a
difference increases with $z$.
\begin{figure}[th]
\epsfxsize=12.98cm \centerline{\epsfbox{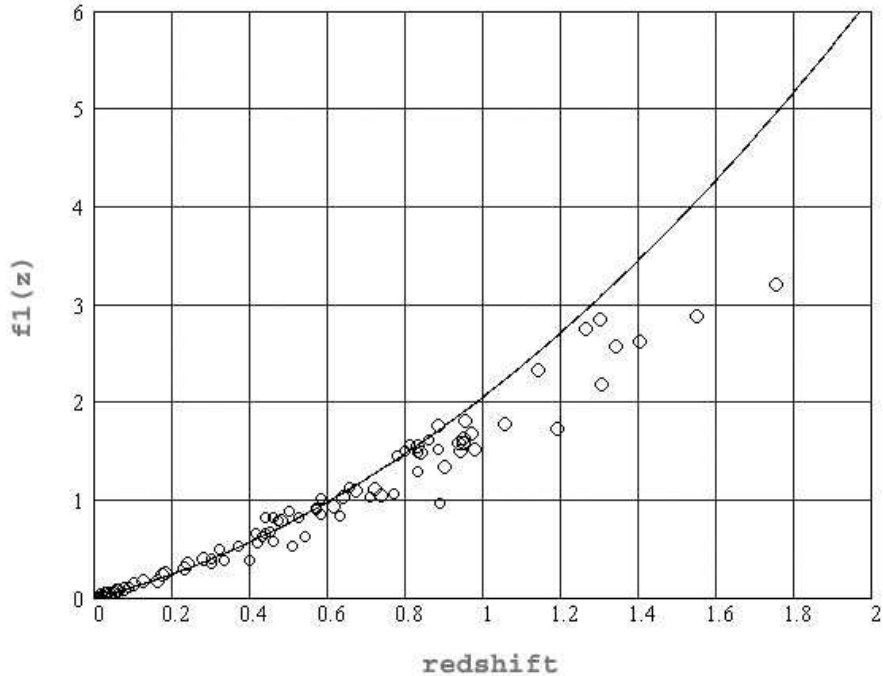}}
\caption{Predicted values of $f_{1}(z)$ (solid line) and
observations (points) from \cite{203} transformed to a linear
scale}
\end{figure}
It would be explained in the model as a result of specific
deformation of SN spectra due to a discrete character of photon
energy losses. Today, a theory of this effect does not exist.

In this model, the Hubble constant may be computed. Let us
consider that a full redshift magnitude is caused by an
interaction with single gravitons, and $\sigma (E,\epsilon)$ is a
cross-section of interaction by forehead collisions of a photon
with an energy $E$ with a graviton, having an energy $\epsilon.$
Let us introduce a new dimensional constant $D$, so that for
forehead collisions:
\begin{equation}
\sigma (E,\epsilon)= D \cdot E \cdot \epsilon.
\end{equation}
Then
\begin{equation}
H= {1 \over 2\pi} D \cdot \bar \epsilon \cdot (\sigma T^{4}),
\end{equation}
where $\bar \epsilon$ is an average graviton energy. Assuming $T
\sim 3 K,~ \bar \epsilon \sim 10^{-4}~ eV,$ and $H = 1.6 \cdot
10^{-18}~ s^{-1},$ we get the following rough estimate for $D:$ $D
\sim 10^{-27}~ m^{2}/eV^{2},$ that gives us the phenomenological
estimate of cross-section by the same and equal $E$ and $\bar
\epsilon$: $\sigma (E,\bar \epsilon) \sim 10^{-35}~ m^{2}.$
\par
It follows from a universality of gravitational interaction, that
not only photons, but all other objects, moving relative to the
background, should lose their energy, too, due to such a quantum
interaction with gravitons. If $a=H/c,$ it turns out that massive
bodies must feel a constant  deceleration of the same order of
magnitude as a small additional acceleration of NASA cosmic probes
(the Pioneer anomaly). We get for the body acceleration $w \equiv
dv/dt$ by a non-zero velocity:
\begin{equation}
w = - ac^{2}(1-v^{2}/c^{2}).
\end{equation}
For small velocities: $w \simeq - Hc.$ If the Hubble constant $H$
is equal to $2.14 \cdot 10^{-18} s^{-1}$ (it is the theoretical
estimate of $H$ in this approach), a modulus of the acceleration
will be equal to $|w|= 6.419 \cdot 10^{-10} \ m/s^{2},$ that has
the same order of magnitude as a value of the observed additional
acceleration  $(8.74 \pm 1.33) \cdot 10^{-10} m/s^2$ for NASA
probes \cite{1}.
\section[3]{Gravity as the screening effect}
It was shown by the author \cite{6,06,006} that screening the
background of super-strong interacting gravitons creates for any
pair of bodies both attraction and repulsion forces due to
pressure of gravitons. For single gravitons, these forces are
approximately balanced, but each of them is much bigger than a
force of Newtonian attraction. If single gravitons are pairing, an
attraction force due to pressure of such graviton pairs is twice
exceeding a corresponding repulsion force if graviton pairs are
destructed by collisions with a body. In such the model, the
Newton constant is connected with the Hubble constant that gives a
possibility to obtain a theoretical estimate of the last. We deal
here with a flat non-expanding universe fulfilled with
super-strong interacting gravitons; it changes the meaning of the
Hubble constant which describes magnitudes of three small effects
of quantum gravity but not any expansion or an age of the
universe.
\subsection[3.1]{Pressure force of single gravitons}
If masses of two bodies are $m_{1}$ and $m_{2}$ (and energies
$E_{1}$ and $E_{2}$), $\sigma (E_{1},\epsilon)$ is a cross-section
of interaction of body $1$ with a graviton with an energy
$\epsilon=\hbar \omega,$ where $\omega$ is a graviton frequency,
$\sigma (E_{2},\epsilon)$ is the same cross-section for body $2.$
Then the following attractive force will act between bodies $1$
and $2:$
\begin{equation}
F_{1}= \int_{0}^{\infty} {\sigma (E_{2},<\epsilon>) \over 4 \pi
r^{2}} \cdot 4 \sigma (E_{1},<\epsilon>)\cdot {1 \over 3} \cdot {4
f(\omega, T) \over c} d\omega.
\end{equation}
If $f(\omega, T)$ is described with the Planck formula, and
$\bar{n} \equiv {1/ (\exp(x)-1)}$ is an average number of
gravitons in a flat wave with a frequency $\omega$ (on one mode of
two distinguishing with a projection of particle spin), $P(n,x)$
is a probability of that in a realization of flat wave a number of
gravitons is equal to $n,$ we shall have for $<\epsilon>$ the
following expression (for more details, see \cite{500}):
\begin{equation}
<\epsilon>= \hbar \omega (1-P(0,x))\bar{n}^{2}\exp(-\bar{n}).
\end{equation}
A quantity $<\epsilon>$ is another average energy of running
gravitons with a frequency $\omega$ taking into account a
probability of that in a realization of flat wave a number of
gravitons may be equal to zero, and that not all of gravitons ride
at a body. Then an attractive force $F_{1}$ will be equal to:
\begin{equation}
F_{1}= {4 \over 3}  {{D^{2} E_{1} E_{2}} \over {\pi r^{2} c}}
\int_{0}^{\infty} {{{\hbar}^{3} \omega^{5}} \over {4\pi^{2}c^{2}}}
(1-P(0,x))^{2}\bar{n}^{5}\exp(-2\bar{n}) d\omega
\end{equation}
$$= {1 \over 3} \cdot {{D^{2} c (kT)^{6} m_{1} m_{2}} \over
{\pi^{3}\hbar^{3}r^{2}}} \cdot I_{1},$$ where $I_{1}= 5.636 \cdot
10^{-3}.$ When $F_{1}\equiv G_{1} \cdot  m_{1}m_{2}/r^{2},$ the
constant $G_{1}$ is equal to:
\begin{equation}
G_{1} \equiv {1 \over 3} \cdot {D^{2} c(kT)^{6} \over
{\pi^{3}\hbar^{3}}} \cdot I_{1}.
\end{equation}
By $T=2.7~ K:$ $G_{1} =1215.4 \cdot G,$ that is three order
greater than the Newton constant, $G.$
\par
But if single gravitons are elastically scattered with body $1,$
then our reasoning may be reversed: the same portion of scattered
gravitons will create a repulsive force $F_{1}^{'}$ acting on body
$2$ and equal to $F_{1}^{'} =F_{1}.$ So, for bodies which
elastically scatter gravitons, screening a flux of single
gravitons does not ensure Newtonian attraction. But for black
holes which absorb any particles and do not re-emit them, we will
have $F_{1}^{'} =0.$ It means that such the object would attract
other bodies with a force which is proportional to $G_{1}$ but not
to $G,$ i.e. Einstein's equivalence principle would be violated
for them. This conclusion stays in force for the case of graviton
pairing, too.
\subsection[3.2]{Graviton pairing}
To ensure an attractive force which is not equal to a repulsive
one, particle correlations should differ for {\it in} and {\it
out} flux. For example, single gravitons of running flux may
associate in pairs \cite{6}. If such pairs are destructed by
collision with a body, then quantities $<\epsilon>$ will be
distinguished for running and scattered particles. Graviton
pairing may be caused with graviton's own gravitational attraction
or gravitonic spin-spin interaction. Left an analysis of the
nature of graviton pairing for the future; let us see that gives
such the pairing.
\par
To find an average number of pairs $\bar{n}_{2}$ in a wave with a
frequency $\omega$ for the state of thermodynamic equilibrium, one
may replace $\hbar \rightarrow 2\hbar$ by deducing the Planck
formula. Then an average number of pairs will be equal to:
\begin{equation}
\bar{n}_{2} ={1 \over {\exp(2x)-1}},
\end{equation}
and an energy of one pair will be equal to $2\hbar \omega.$ It is
important that graviton pairing does not change a number of
stationary waves, so as pairs nucleate from existing gravitons.
The question arises: how many different modes, i.e. spin
projections, may graviton pairs have? We assume here that the
background of initial gravitons consists of two modes. For
massless transverse bosons, it takes place as by spin $1$ as by
spin $2.$ If graviton pairs have maximum spin $2,$ then single
gravitons should have spin $1.$ But from such particles one may
constitute four combinations: $\uparrow \uparrow, \ \downarrow
\downarrow $ (with total spin $2$), and $\uparrow \downarrow, \
\downarrow\uparrow$ (with total spin $0).$ All these four
combinations will be equiprobable if spin projections $\uparrow$
and $\downarrow$ are equiprobable in a flat wave (without taking
into account a probable spin-spin interaction).
\par
But it follows from the energy conservation law that composite
gravitons should be distributed only in two modes. So as
\begin{equation}
\lim_{x \to 0} {\bar{n}_{2} \over \bar{n}} ={1/2},
\end{equation}
then by $x \rightarrow 0$ we have $2\bar{n}_{2}=\bar{n},$ i.e. all
of gravitons are pairing by low frequencies. An average energy on
every mode of pairing gravitons is equal to $2 \hbar \omega
\bar{n}_{2},$ the one on every mode of single gravitons - to
$\hbar \omega \bar{n}.$ These energies are equal by $x \rightarrow
0,$ because of that, the numbers of modes are equal, too, if the
background is in the thermodynamic equilibrium with surrounding
bodies. The above reasoning does not allow to choose a spin value
$2$ or $0$ for composite gravitons. A choice of namely spin $2$
would ensure the following proposition: all of gravitons in one
realization of flat wave have the same spin projections. From
another side, a spin-spin interaction would cause it.
\par
The spectrum of composite gravitons is also the Planckian one, but
with a smaller temperature $T_{2} \equiv (1/8)^{1/4}T = 0.5946 \
T.$
\par
It is important that the graviton pairing effect does not change
computed values of the Hubble constant and of anomalous
deceleration of massive bodies: twice decreasing of a sub-system
particle number due to the pairing effect is compensated with
twice increasing the cross-section of interaction of a photon or
any body with such the composite gravitons. Non-pairing gravitons
with spin $1$ give also its contribution in values of redshifts,
an additional relaxation of light intensity due to non-forehead
collisions with gravitons, and  anomalous deceleration of massive
bodies moving relative to the background.

\subsection[3.3]{Computation of the Newton constant, and a connection
between the two fundamental constants, $G$ and $H$ }  If running
graviton pairs ensure for two bodies an attractive force $F_{2},$
then a repulsive force due to re-emission of gravitons of a pair
alone will be equal to $F_{2}^{'} =F_{2}/2.$ It follows from that
the cross-section for {\it single additional scattered} gravitons
of destructed pairs will be twice smaller than for pairs
themselves (the leading factor $2\hbar \omega$ for pairs should be
replaced with $\hbar \omega$ for single gravitons). For pairs, we
introduce here the cross-section $ \sigma (E_{2},<\epsilon_{2}>),$
where $<\epsilon_{2}>$ is an average pair energy with taking into
account a probability of that in a realization of flat wave a
number of graviton pairs may be equal to zero, and that not all of
graviton pairs ride at a body ($<\epsilon_{2}>$ is an analog of
$<\epsilon>$). Replacing $\bar{n} \rightarrow \bar{n}_{2},~ \hbar
\omega \rightarrow 2\hbar \omega,$ and $P(n,x) \rightarrow
P(n,2x),$ where $P(0,2x)= \exp(-\bar{n}_{2}),$ we get for graviton
pairs:
\begin{equation}
<\epsilon_{2}> \sim 2\hbar \omega
(1-P(0,2x))\bar{n}_{2}^{2}\exp(-\bar{n}_{2}).
\end{equation}
This expression does not take into account only that beside pairs
there may be single gravitons in a realization of flat wave. To
reject cases when, instead of a pair, a single graviton runs
against a body (a contribution of such gravitons in attraction and
repulsion is the same), we add the factor $P(0,x)$ into
$<\epsilon_{2}>:$
\begin{equation}
<\epsilon_{2}> = 2\hbar \omega
(1-P(0,2x))\bar{n}_{2}^{2}\exp(-\bar{n}_{2}) \cdot P(0,x).
\end{equation}
Then a force of attraction of two bodies due to pressure of
graviton pairs, $F_{2}$, - in the full analogy with (19) - will be
equal to\footnote{In initial version of this paper, factor 2 was
lost in the right part of Eq. (15), and the theoretical values of
$D$ and $H$ were overestimated of $\sqrt{2}$ times}:
\begin{equation}
F_{2}= \int_{0}^{\infty} {\sigma (E_{2},<\epsilon_{2}>) \over 4
\pi r^{2}} \cdot 4 \sigma (E_{1},<\epsilon_{2}>)\cdot {1 \over 3}
\cdot {4 f_{2}(2\omega,T) \over c} d\omega =
\end{equation}
$$ {8 \over 3} \cdot
{D^{2} c(kT)^{6} m_{1}m_{2} \over {\pi^{3}\hbar^{3}r^{2}}}\cdot
I_{2},$$ where $I_{2} = 2.3184 \cdot 10^{-6}.$ The difference $F$
between attractive and repulsive forces will be equal to:
\begin{equation}
F \equiv F_{2}- F_{2}^{'}={1 \over 2}F_{2} \equiv G_{2}{m_{1}m_{2}
\over r^{2}},
\end{equation}
where the constant $G_{2}$ is equal to:
\begin{equation}
G_{2} \equiv {4 \over 3} \cdot {D^{2} c(kT)^{6} \over
{\pi^{3}\hbar^{3}}} \cdot I_{2}.
\end{equation}
Both $G_{1}$ and $G_{2}$ are proportional to $T^{6}$ (and $H \sim
T^{5},$ so as $\bar{\epsilon} \sim T$).
\par
If one assumes that $G_{2}=G,$ then it follows from (17) that by
$T=2.7K$ the constant $D$ should have the value: $D=0.795 \cdot
10^{-27}{m^{2} / eV^{2}}.$
\par
We can use (5) and (17) to establish a connection between the two
fundamental constants, $G$ and $H$, under the condition that
$G_{2}=G.$ We have for $D:$
\begin{equation}
D= {2\pi H \over \bar{\epsilon} \sigma T^{4}}= {2 \pi^{5} H \over
15 k \sigma T^{5} I_{4}};
\end{equation}
then
\begin{equation}
G=G_{2} = {4 \over 3} \cdot {D^{2} c(kT)^{6} \over
{\pi^{3}\hbar^{3}}} \cdot I_{2}= \\
{64 \pi^{5} \over 45} \cdot {H^{2}c^{3}I_{2} \over \sigma T^{4}
I_{4}^{2}}.
\end{equation}
So as the value of $G$ is known much better than the value of $H,$
let us express $H$ via $G:$
\begin{equation}
H= (G  {45 \over 64 \pi^{5}}  {\sigma T^{4} I_{4}^{2} \over
{c^{3}I_{2}}})^{1/2}= 2.14 \cdot 10^{-18}~s^{-1},
\end{equation}
or in the units which are more familiar for many of us: $H=66.875
\ km \cdot s^{-1} \cdot Mpc^{-1}.$
\par
This value of $H$ is in the good accordance with the majority of
present astrophysical estimations \cite{2,512,513} (for example,
the estimate $(72 \pm 8)$ km/s/Mpc has been got from SN1a
cosmological distance determinations in \cite{513}), but it is
lesser than some of them \cite{512a} and than it follows from the
observed value of anomalous acceleration of Pioneer 10 \cite{1}.

\subsection[3.4]{Restrictions on a geometrical language in gravity}
The described quantum mechanism of classical gravity gives
Newton's law with the constant $G_{2}$ value (17) and the
connection (19) for the constants $G_{2}$ and $H.$  We have
obtained the rational value of $H$ (20) by $G_{2} = G,$ if the
condition of big distances is fulfilled:
\begin{equation}
\sigma (E_{2},<\epsilon>) \ll 4 \pi r^{2}.
\end{equation}
Because it is known from experience that for big bodies of the
solar system, Newton's law is a very good approximation, one would
expect that this condition is fulfilled, for example, for the pair
Sun-Earth. But assuming $r=1 \ AU$ and $E_{2}=m_{\odot}c^{2},$ we
obtain assuming for rough estimation $<\epsilon> \rightarrow
\bar{\epsilon}:$ $${\sigma (E_{2},<\epsilon>) \over 4 \pi r^{2}}
\sim 4 \cdot 10^{12}. $$ It means that in the case of interaction
of gravitons or graviton pairs with the Sun in the aggregate, the
considered quantum mechanism of classical gravity could not lead
to Newton's law as a good approximation. This "contradiction" with
experience is eliminated if one assumes that gravitons interact
with "small particles" of matter - for example, with atoms. If the
Sun contains of $N$ atoms, then $\sigma (E_{2},<\epsilon>)=N
\sigma (E_{a},<\epsilon>),$ where $E_{a}$ is an average energy of
one atom. For rough estimation we assume here that $E_{a}=E_{p},$
where $E_{p}$ is a proton rest energy; then it is $N \sim
10^{57},$ i.e. ${\sigma (E_{a},<\epsilon>)/ 4 \pi r^{2}} \sim
10^{-45} \ll 1.$
\par
This necessity of "atomic structure" of matter for working the
described quantum mechanism is natural relative to usual bodies.
But would one expect that black holes have a similar structure? If
any radiation cannot be emitted with a black hole, a black hole
should interact with gravitons as an aggregated object, i.e. this
condition for a black hole of sun mass has not been fulfilled even
at distances $\sim 10^{6} \ AU.$
\par
For bodies without an atomic structure, the allowances, which are
proportional to $D^{3}/ r^{4}$ and are caused by decreasing a
gravitonic flux due to the screening effect, will have a factor
$m_{1}^{2}m_{2}$ or $m_{1}m_{2}^{2}.$ These allowances break the
equivalence principle for such the bodies.
\par
For bodies with an atomic structure, a force of interaction is
added up from small forces of interaction of their "atoms": $$ F
\sim N_{1}N_{2}m_{a}^{2}/r^{2}=m_{1}m_{2}/r^{2},$$ where $N_{1}$
and $N_{2}$ are numbers of atoms for bodies $1$ and $2$. The
allowances to full forces due to the screening effect will be
proportional to the quantity: $N_{1}N_{2}m_{a}^{3}/r^{4},$ which
can be expressed via the full masses of bodies as
$m_{1}^{2}m_{2}/r^{4}N_{1}$ or $m_{1}m_{2}^{2}/r^{4}N_{2}.$ By big
numbers $N_{1}$ and $N_{2}$ the allowances will be small. The
allowance to the force $F,$ acting on body $2,$ will be equal to:
\begin{equation}
\Delta F ={1 \over 2 N_{2}} \int_{0}^{\infty} {\sigma^{2}
(E_{2},<\epsilon_{2}>) \over (4 \pi r^{2})^{2}} \cdot 4 \sigma
(E_{1},<\epsilon_{2}>)\cdot {1 \over 3} \cdot {4 f_{2}(2\omega,T)
\over c} d\omega
\end{equation}
$$={2 \over 3N_{2}} \cdot {{D^{3} c^{3} (kT)^{7}
m_{1} m_{2}^{2}} \over {\pi^{4}\hbar^{3}r^{4}}} \cdot I_{3},$$
(for body $1$ we shall have the similar expression if replace
$N_{2} \rightarrow N_{1}, \ m_{1}m_{2}^{2} \rightarrow
m_{1}^{2}m_{2}$), where $ I_{3} = 1.0988 \cdot 10^{-7}. $
\par
Let us find the ratio:
\begin{equation}
{\Delta F \over F} = {D E_{2} kT \over {N_{2} 2\pi r^{2}}} \cdot
{I_{3} \over I_{2}}.
\end{equation}
Using this formula, we can find by $E_{2}=E_{\odot}, \ r=1 \ AU:$
\begin{equation}
{\Delta F \over F} \sim 10^{-46}.
\end{equation}
\par
An analogical allowance to the force $F_{1}$ has by the same
conditions the order $\sim 10^{-48}F_{1},$ or $\sim 10^{-45}F.$
One can replace $E_{p}$ with a rest energy of very big atom - the
geometrical approach will left a very good language to describe
the solar system. We see that for bodies with an atomic structure
the considered mechanism leads to very small deviations from
Einstein's equivalence principle, if the condition of big
distances is fulfilled for microparticles, which prompt interact
with gravitons.
\par
For small distances we shall have:
\begin{equation}
\sigma (E_{2},<\epsilon>) \sim 4 \pi r^{2}.
\end{equation}
It takes place by $E_{a}=E_{p}, \ <\epsilon> \sim 10^{-3} \ eV$
for $r \sim 10^{-11} \ m.$ This quantity is many orders larger
than the Planck length. The equivalence principle should be broken
at such distances.

\section[4]{Some cosmological consequences of the model}
If the described model of redshifts is true, what is a picture of
the universe? It is interesting that in a frame of this model,
every observer has two own spheres of observability in the
universe (two different cosmological horizons exist for any
observer) \cite{44,55}. One of them is defined by maximum existing
temperatures of remote sources - by big enough distances, all of
them will be masked with the CMB radiation. Another, and much
smaller, sphere depends on their maximum luminosity - the
luminosity distance increases with a redshift much quickly than
the geometrical one. The ratio of the luminosity distance to the
geometrical one is the quickly increasing function of $z:$
$D_{L}(z)/r(z)= (1+z)^{(1+b)/2},$ which does not depend on the
Hubble constant.  An outer part of the universe will drown in a
darkness. We can assume that the graviton background and the
cosmic microwave one are in a state of thermodynamical
equilibrium, and have the same temperatures. CMB itself may arise
as a result of cooling any light radiation up to reaching this
equilibrium. Then it needs $z \sim 1000$ to get through the very
edge of our cosmic "ecumene". Some other possible cosmological
consequences of an existence of the graviton background were
described in \cite{08,6}.
\par
The graviton background may turn up "a perpetual engine" of the
universe, pumping energy from any radiation to massive objects. An
equilibrium state of the background will be ensured by such a
temperature $T,$ for which an energy profit of the background due
to an influx of energy from radiation will be equal to a loss of
its energy due to a catch of virtual massive gravitons with "black
holes" or other massive objects. In such the picture, the chances
are that "black holes" would turn out "germs" of galaxies. After
accumulation of a big enough energy by a "black hole" (to be more
exact, by a super-compact massive object) by means of a catch of
virtual massive gravitons, the one would be absolved from an
energy excess in via ejection of matter, from which stars of
galaxy should form. It awaits to understand else in such the
approach how usual matter particles form from virtual massive
gravitons. \par There is a very interesting but non-researched
possibility: due to relative decreasing of an intensity of
graviton pair flux in an internal area of galaxies (pairs are
destructed under collisions with matter particles), the effective
Newton constant may turn out to be running on galactic scales. It
might lead to something like to the modified Newtonian dynamics
(MOND) by  Mordehai Milgrom (about MOND, for example, see
\cite{99}). But to evaluate this effect, one should take into
account a relaxation process for pairs, about which we know
nothing today. It is obvious only that gravity should be stronger
on a galactic periphery. The renormalization group approach to
gravity leads to modifications of the theory of general relativity
on galactic scales \cite{799,899}, and a growth of Newton's
constant at large distances takes place, too. Kepler's third law
receives quantum corrections that may explain the flat rotation
curves of the galaxies.

\section[5]{How to verify the main conjecture of this approach in
a laser experiment on the Earth} I would like to show here (see
\cite{112,500}) a full realizability at present time of verifying
my basic conjecture about the quantum gravitational nature of
redshifts in a ground-based laser experiment. Of course, many
details of this precision experiment will be in full authority of
experimentalists. \par It was not clear in 1995 how big is a
temperature of the graviton background, and my proposal \cite{111}
to verify the conjecture about the described local quantum
character of redshifts turned out to be very rigid: a laser with
instability of $\sim 10^{-17}$ hasn't appeared after 10 years. But
if $T=2.7 K$, the satellite of main laser line of frequency $\nu$
after passing the delay line will be red-shifted at $\sim 10^{-3}$
eV/h and its position will be fixed (see Fig. 2). It will be
\begin{figure}[th]
\centerline{\includegraphics[width=12.98cm]{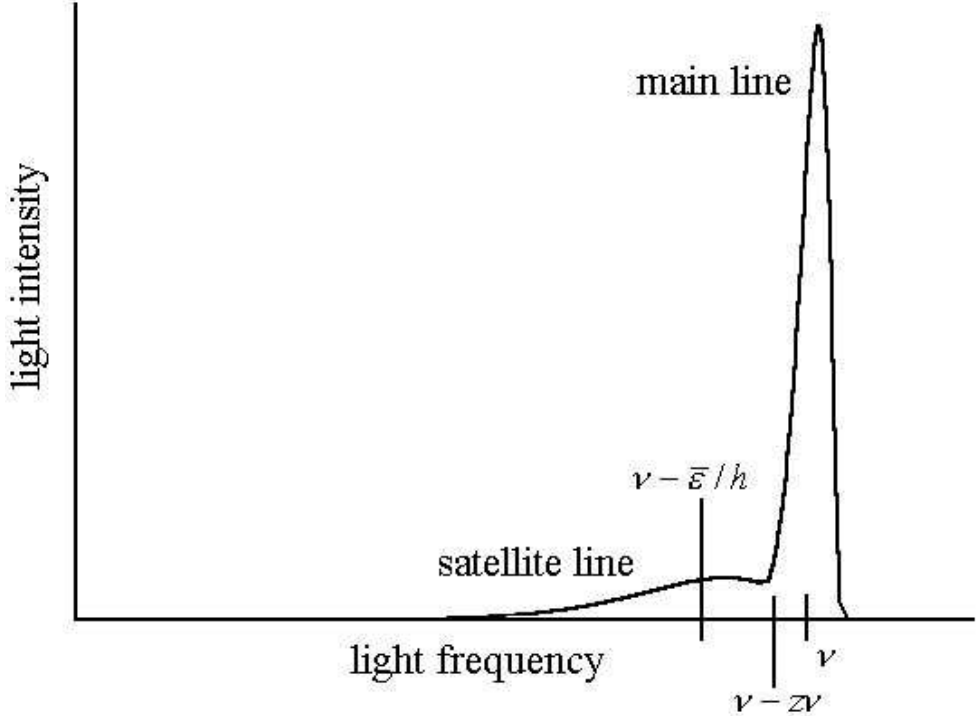}}
\caption{The main line and the expected red-shifted satellite line
of a stable laser radiation spectrum after a delay line.
Satellite's position should be fixed near $\nu -\bar{\epsilon}/h$,
and its intensity should linear rise with a path of photons in a
delay line, $l$. A center-of-mass of both lines is expected to be
approximately near $\nu - z \nu$.}
\end{figure}
caused by the fact that on a very small way in the delay line only
a small part of photons may collide with gravitons of the
background. The rest of them will have unchanged energies. The
center-of-mass of laser radiation spectrum should be shifted
proportionally to a photon path. Then due to the quantum nature of
shifting process, the ratio of satellite's intensity to main
line's intensity should have the order: $\sim {h\nu \over
\bar{\epsilon}}{H\over c} l,$ where $l$ is a path of laser photons
in a vacuum tube of delay line. It gives us a possibility to plan
a laser-based experiment to verify the basic conjecture of this
approach with much softer demands to the equipment. An instability
of a laser of a power $P$ must be only $\ll 10^{-3}$ if a photon
energy is of $\sim 1~eV$. It will be necessary to compare
intensities of the red-shifted satellite at the very beginning of
the path $l$ and after it. Given a very low signal-to-noise ratio,
one could use a single photon counter to measure the intensities.
When $q$ is a quantum output of a cathode of the used
photomultiplier (a number of photoelectrons is $q$ times smaller
than a number of photons falling to the cathode), $N_{n}$ is a
frequency of its noise pulses, and $n$ is a desired ratio of a
signal to noise's standard deviation, then an evaluated time
duration $t$ of data acquisition would have the order:
\begin{equation}
t= {\bar{\epsilon}^{2}c^{2} \over H^{2}} {n^{2}N_{n} \over q^{2}
P^{2} l^{2} }.
\end{equation}
Assuming $n=10,~N_{n}=10^{3}~s^{-1},~ q=0.3, ~P=100~ mW,~ l=100
~m, $ we would have the estimate: $t= 200,000 $ years, that is
unacceptable. But given $P=300~W$, we get: $t \sim 8$ days, that
is acceptable for the experiment of such the potential importance.
Of course, one will rather choose a bigger value of $l$ by a small
laser power forcing a laser beam to whipsaw many times between
mirrors in a delay line - it is a challenge for experimentalists.
\section[6]{Gravity in a frame of non-linear and non-local QED? -
the question only to the Nature} From thermodynamic reasons, it is
assumed here that the graviton background has the same temperature
as the microwave background. Also it follows from the condition of
detail equilibrium, that both backgrounds should have the
Planckian spectra. Composite gravitons will have spin $2$, if
single gravitons have the same spin as photons. The question
arise, of course: how are gravitons and photons connected? Has the
conjecture by Adler et al. \cite{a98,a99} (that a graviton with
spin $2$ is composed with two photons) chances to be true?
Intuitive demur calls forth a huge self-action, photons should be
endued with which if one unifies the main conjecture of this
approach with the one by Adler et al. - but one may get a unified
theory on this way.
\par To verify this combined conjecture in experiment, one would
search for transitions in interstellar gas molecules caused by the
microwave background, with an angular momentum change
corresponding to absorption of spin $2$ particles (photon pairs).
A frequency of such the transitions should correspond to an
equivalent temperature  of the sub-system of these composite
particles $T_{2}=0.5946~ T,$ if $T$ is a temperature  of the
microwave background.
\par
From another side, one might check this conjecture in a laser
experiment, too (see \cite{204,500}). Taking two lasers with
photon energies $h\nu_{1}$ and $h\nu_{2}$, one may force laser
beams to collide on a way $L$ (see Fig. 3). If photons are
self-interacting particles, we might wait that photons with
energies $h\nu_{1}-h\nu_{2}$, if $h\nu_{1}
> h\nu_{2}$, would arise after collisions of initial photons.
\begin{figure}[th]
\centerline{\includegraphics[width=12.98cm]{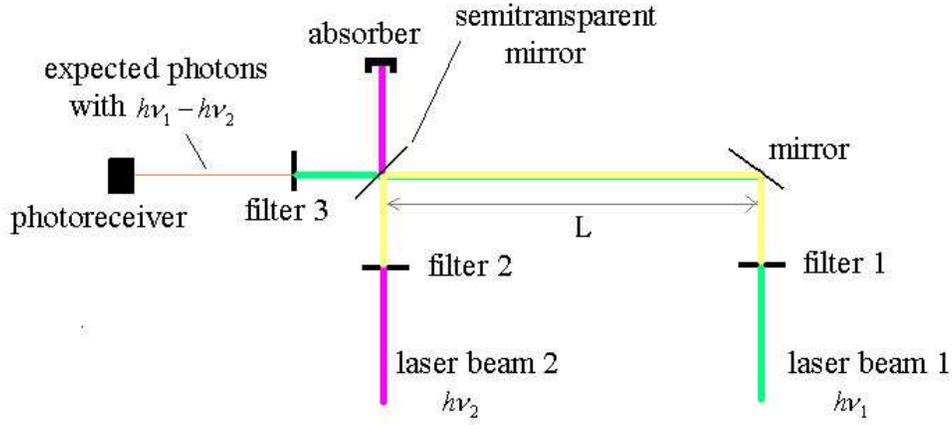}}
\caption{The scheme of laser beam passes. Two laser beams 1 and 2
collide into the area with a length $L$. An expected beam of
photons with energies $h\nu_{1}-h\nu_{2}$ falls to a
photoreceiver.}
\end{figure}
If we {\it assume (only here)} that single gravitons are identical
to photons, it will be necessary to take into account the
following circumstances to calculate an analog of the Hubble
constant for this experiment: an average graviton energy should be
replaced with $h\nu_{2}$, the factor $1/2\pi$ in (5) should be
replaced with $1/\varphi$, where $\varphi$ is a divergence of
laser beam 2, and one must use a quantity $P/S$ instead of $\sigma
T^{4}$ in (5), where $P$ is a laser 2 power and $S$ is a
cross-section of its beam. Together all it means that we should
replace the Hubble constant with its analog for a laser beam
collision, $H_{laser}$:
\begin{equation}
H \rightarrow H_{laser} = {1 \over \varphi} \cdot D \cdot
h\nu_{2}\cdot {P \over S}.
\end{equation}
Taken $\varphi=10^{-4}$, $h\nu_{2} \sim 1~eV$, $P \sim 10~mW$, and
$P/S \sim 10^{3}~W/m^{2}$, that is characterizing a He-Ne laser,
we get the estimate: $H_{laser} \sim 0.06 ~s^{-1}$. Then photons
with energies $h\nu_{1}-h\nu_{2}$ would fall to a photoreceiver
with a frequency which should linearly rise with $L$
(proportionally to ${H_{laser} \over c} \cdot L$), and it would be
of $10^{7}~s^{-1}$ if both lasers have equal powers $\sim 10~mW$,
and $L\sim 1~m$. It is a big enough frequency to give us a
possibility to detect easy a flux of these expected photons in IR
band.
\par If this tentative non-linear vacuum effect exists, it would
lead us far beyond standard quantum electrodynamics to take into
account new non-linearities (which are not connected with the
electron-positron pair creation) and an essential impact of such a
non-locally born object as the graviton background.

\section[7]{Conclusion}
It follows from the above consideration that the geometrical
description of gravity should be a good idealization for any pair
of bodies at a big distance by the condition of an "atomic
structure" of matter. This condition cannot be accepted only for
black holes which must interact with gravitons as aggregated
objects. In addition, the equivalence principle is roughly broken
for black holes, if the described quantum mechanism of classical
gravity is realized in the nature. Because attracting bodies are
not initial sources of gravitons, a future theory must be
non-local in this sense to describe gravitons running from
infinity. The described quantum mechanism of classical gravity is
obviously asymmetric relative to the time inversion. By the time
inversion, single gravitons would run against bodies to form pairs
after collisions with bodies. It would lead to replacing a body
attraction with a repulsion. But such the change will do
impossible the graviton pairing. Cosmological models with the
inversion of the time arrow were considered by Sakharov
\cite{a16}. Penrose has noted that a hidden physical law may
determine the time arrow direction \cite{a17}; it will be very
interesting if namely realization in the nature of Newton's law
determines this direction. \par A future theory dealing with
gravitons as usual particles should have a number of features
which are not characterizing any existing model to image the
considered here features of the possible quantum mechanism of
gravity. If this mechanism is realized in the nature, both the
general relativity and quantum mechanics should be modified. Any
divergencies, perhaps, would be not possible in such the model
because of natural smooth cut-offs of the graviton spectrum from
both sides. Gravity at short distances, which are much bigger than
the Planck length, needs to be described only in some unified
manner.


\begin{thebibliography}{References                        }
\bibitem{1}
Anderson, J.D. et al. {\it Phys. Rev. Lett.} 1998, {\it 81,} 2858;
{\it Phys. Rev.} 2002, {\it D65,} 082004; [gr-qc/0104064 v4].
\bibitem{2}
Riess, A.G. et al. {\it AJ} 1998, {\it 116,} 1009.
\bibitem{3}
Perlmutter, S. et al. {\it ApJ} 1999, {\it 517,} 565.
\bibitem{4}
Nesvizhevsky, V.V. et al. {\it Nature} 2002, {\it 415,} 297.
\bibitem{5}
Ivanov, M.A. {\it General Relativity and Gravitation} 2001, {\it
33,} 479; Erratum: 2003, {\it 35,} 939; [astro-ph/0005084 v2].
\bibitem{500}
Ivanov, M.A. Gravitons as super-strong interacting particles, and
low-energy quantum gravity [hep-th/0506189],
[http://ivanovma.narod.ru/nova04.pdf].
\bibitem{5a}
Ivanov, M.A. [gr-qc/0009043]; Proc. of the Int. Symp. "Frontiers
of Fundamental Physics 4" (9-13 Dec 2000, Hyderabad, India),
Sidharth, B.G., Altaisky, M.V., Eds.; Kluwer Academic/Plenum
Publishers: August 2001; Proc. of the 4th Edoardo Amaldi
Conference on Gravitational Waves (Perth, Western Australia, 8-13
July 2001) {\it Class. Quantum Grav.} 2002, {\it 19,} 1351.
\bibitem{6}
Ivanov, M.A. Screening the graviton background, graviton pairing,
and Newtonian gravity [gr-qc/0207006].
\bibitem{213}
Ivanov, M.A. Another origin of cosmological redshifts
[astro-ph-0405083].
\bibitem{203}
Riess, A.G. et al. Type Ia Supernova Discoveries at $z > 1$ From
the Hubble Space Telescope ... [astro-ph/0402512] (to appear in
{\it ApJ,} 2004).
\bibitem{111}
Ivanov, M.A. Contribution to the Quantum Electronics and Laser
Science Conference (QELS'95), May 21-26, 1995, Baltimore, USA;
paper number: QThG1.
\bibitem{204}
Ivanov, M.A. Contribution to The sixth international symposium
"Frontiers of Fundamental and Computational Physics" (FFP6), 26-29
September 2004, Udine, Italy; [astro-ph/0409631].
\bibitem{06}
Ivanov, M.A. Contribution to The Tenth Marcel Grossmann Meeting
(MG10), 20-26 July 2003, Rio de Janeiro, Brazil; [gr-qc/0307093].
\bibitem{006}
Ivanov, M.A. Contribution to the Conference "Thinking, Observing
and Mining the Universe" (Thinking2003), 22-27 Sep 2003, Sorrento,
Italy; [astro-ph/0309566].
\bibitem{512}
Freedman, W. L. et al. {\it ApJ} 2001, {\it 553,} 47.
\bibitem{513}
Filippenko, A.V. [astro-ph/0410609]; to be published in {\it White
Dwarfs: Probes of Galactic Structure and Cosmology,}  Sion, E. M.,
Shipman, H. L., Vennes, S., Eds.; Kluwer: Dordrecht.
\bibitem{512a}
Willick, J.A., Puneet Batra. {\it ApJ} 2001, {\it 548,} 564.
\bibitem{44}
Ivanov, M.A. In {\it Searches for a mechanism of gravity,} Ivanov,
M.A., Savrov, L.A., Eds.; Nikolaev, Yu.A. Publisher: Nizhny
Novgorod, 2004, pp 266-273 (in Russian).
\bibitem{55}
Ivanov, M.A. A quantum gravitational model of redshifts
[astro-ph/0409111].
\bibitem{08}
Ivanov, M.A. Model of graviton-dusty universe [gr-qc/0107047].
\bibitem{99}
Milgrom, M. In the Proceedings of the II Int. Workshop on the
Identification of Dark Matter, Buxton, England, 1998; World
Scientific: Singapore, 1999; [astro-ph/9810302].
\bibitem{799}
Shapiro, I.L., Sola, J., Stefancic, H. Running G and $\Lambda$ at
low energies from physics at $M_{X}$: possible cosmological and
astrophysical implications [hep-ph/0410095].
\bibitem{899}
Reuter, M., Weyer, H. Running Newton Constant, Improved
Gravitational Actions, and Galaxy Rotation Curves
[hep-th/0410117].
\bibitem{112}
Ivanov, M.A. Contribution to The 14th Workshop on General
Relativity and Gravitation (JGRG14), Nov 29 - Dec 3 2004, Kyoto,
Japan; [gr-qc/0410076].
\bibitem{a98}
Adler, S.L. et al. {\it Phys. Rev.} 1976, {\it D14,} 359.
\bibitem{a99}
Adler, S.L. {\it Phys. Rev.} 1976, {\it D14,} 379.
\bibitem{a16}
Sakharov, A.D. {\it JETP} 1980, {\it 79,} 689.
\bibitem{a17}
Penrose. R.~ In the Einstein Survey {\it General Relativity,}
Hawking, S.W., Israel, W., Eds.; Cambridge University Press: 1979.
\end{thebibliography}
\end{document}